# Model Visualization in understanding rapid growth of a journal in an emerging area

Snehanshu Saha[1], Poulami Sarkar[1], Archana Mathur[2], Suryoday Basak[1]

[1] Department of Computer Science and Engineering, PESIT South Campus, Bangalore

snehanshusaha@pes.edu, poulamirulz@gmail.com, suryodaybasak@gmail.com

[2] SSIU, Indian Statistical Institute, Mysore Road, Bangalore

mathurarchana77@gmail.com

**ABSTRACT**
A recent independent study resulted in a ranking system which ranked Astronomy and Computing (ASCOM) much higher than most of the older journals highlighting the niche prominence of the particular journal. We investigate the remarkable ascendancy in reputation of ASCOM by proposing a novel differential equation based modeling. The modeling is a consequence of knowledge discovery from big data-centric methods, namely L1-SVD. The inadequacy of the ranking method in explaining the reason behind the growth in reputation of ASCOM is reasonable to understand given that the study was post-facto. Thus, we propose a growth model by accounting for the behavior of parameters that contributes to the growth of a field. It is worthwhile to spend some time in analyzing the cause and control variables behind rapid rise in reputation of a journal in a niche area. We intend to probe and bring out parameters responsible for its growing influence. Delay differential equations are used to model the change of influence on a journal's status by exploiting the effects of historical data.

## INTRODUCTION

With emergence of new areas in the field of interdisciplinary Science and computing, new journals have come into existence all over the world. With the rapid growth in technology, it has become important to compute the pattern in growth [1],[2], [3], [4], [5] of emerging areas. Understanding the growth of an area enables us to estimate influence of a field in the technological world. A domain's growth can be assessed by analyzing journals that have emerged in that domain, not just their count but also the reputation earned. A newly evolved peer-reviewed journal, Astronomy and Computing (ASCOM) publishes in the field where astronomy, computer science and modelling intersect. ASCOM, though newly arrived, has earned good standing and currently boasts of reasonable decent Impact Factor, SNIP and SJR scores when compared with the older and seasoned journals in the same domain. In this project, we aim to understand the pattern in which ASCOM's reputation has grown from its inception in 2013. In an independent study [7,8], we have presented a novel method to compute the rank of ASCOM and found that it is within the top 50 journals in Astronomy and Astrophysics and more importantly, is ranked higher than many of its elders.

We investigate the remarkable ascendancy in reputation of ASCOM by proposing a novel differential equation based modeling in the subsequent sections of the manuscript. Knowledge discovery from big data centric methods, namely L1-SVD [7, 8] serve as the primary motivation for the modeling approach. The l1-SVD approach, though efficient and elegant doesn't explain the reason behind the growth in reputation of ASCOM, a relatively new journal serving fairly niche scientific interests.

**The MODEL:**

We propose a hypothesis: The rate of change of the influence of a journal depends on the influence at current as well as historic time. The hypothesis is represented in the form of equation (1) wherein, delay differential equation has been employed to model the influence of ASCOM. Delay Differential Equations belong to the class of Partial Differential Equations. These are used by the scientific community for modeling dynamic systems for many of the obvious advantages. These equations describe a rate of change of a function, at time t as a function of earlier times. A DDE in its general form can be given by:

$$p'(t) = f(p(t), p(t -\tau)) \qquad (1)$$

Assuming a constant delay of $\tau$.

The equation is given by:

The scibase and scientometric modeling effort is endorsed and supported by IEEE Computer Society Bangalore Chapter.

$$p'(t) = a*p(t) + b*p(-t) \quad (2)$$

Where, p'(t) is the rate of change of a journal's influence over time
p(t) is the influence of journal at time t
p(-t) is the influence at time –t

Here, t represents current time, p(-t) represents influence in the past and p(t) is the current influence. The change of influence depends on both the current, as well as, the historic values of influence.

**Sketch of the solution technique:**

Under such assumptions of steady growth: If we consider a simple model:

$$a*p'(t) = b + c*p(t) \quad \text{where} \quad p(0)=c;$$

$$p'(t) = (b/a)+(c/a)*p(t)$$

Differentiating the above equation with respect to 't' and assuming symmetric influence i.e. p(t)= p(-t) (implying influence patterns are numerically proximal in time series), we obtain

$$p''(t)=(c/2a)[(b/a)+(c/2a)*p(t)+(c/2a)*p(-t)]-(c/2a)[(b/a)+(c/2a)*p(-t)+(c/2a)*p(t)] \quad (3)$$

Case 1: If we assume that there is a symmetry in the journal's influence i.e. p(t)=p(-t) , then p(t) turns out to be a linear function. The trend is shown in Figure 1: The growth of the journal thus is linear.

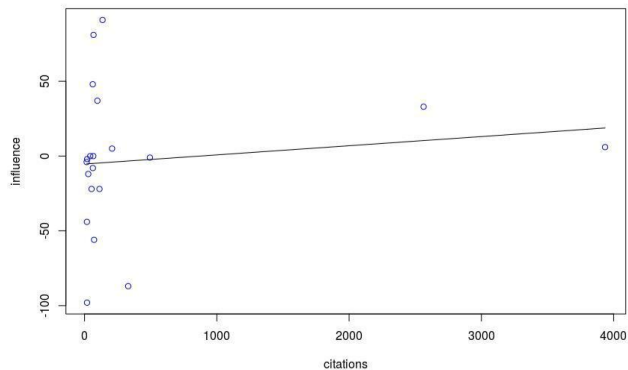

Fig 1: Linear model obtained when we assume p(t)=p(-t)
: The growth is more or less expected and modeling has a little role to play

However the rapid growth in influence and reputation of ASCOM in a very short duration of time does not support the linear growth doctrine. Hence, it is important to investigate the reasons and conditions behind the non-linear growth of the journal's influence. If the assumption on symmetry (i.e. p(t) = p(-t)) in the influence model is not considered (which is a strong assumption and therefore may not be realistic), we would either have an exponential growth or decay trend depending on the coefficients in the model equation. The following derivation considers non-linear growth (influence).

Without assuming a symmetric influence function (which is too strong a condition anyway) in our model, a special case of DDE where the rate of change of influence is expressed both in terms of the current as well as the historical data (in terms of journal's influence) is given by:

$$p''(t) = a*p'(-t)+b*p'(t)$$

$$p''(t) = (a)*-a*p(t)-b*p(-t)] + (b)[(a*p(-t)+b*p(t)]$$

$$p''(t)=(b^2-a^2)p(t), \text{ where } r= \sqrt{(b^2-a^2)}$$
$$p''(t)=r^2*p(t)$$

A classic solution to the second order differential equation is given by:

$$p(t)=A*e^{rt} + B*e^{-rt} .$$

Using the initial conditions, we easily obtain the expression for influence as:

$$p(t)=(c/2*r)*(r+a+b)e^{rt}+(c/2*r)*(r-a-b)*e^{-rt} \quad (4)$$

Re-written as

$$p(t)=w_1*e^{rt} + w_2*e^{-rt} \quad (5)$$
where $w_1 = (c/2r)*(r+a+b)$, $w_2 = (c/2r)*(r-a-b)$.

**Model Visualization:**

The above solution gives rise to three possibilities:

1. r=0: This leads to linear growth (influence)
2. r>0 (b>a) : exponential real solutions
3. r<0: Imaginary solution

We investigate the visual implications [6] below.

Case 1: Linear growth is not relevant in the context of the journal under study,

The scibase and scientometric modeling effort is endorsed and supported by IEEE Computer Society Bangalore Chapter.

Case 2: r>0 (b>a): exponential real solutions when the parameters a and b are varied.

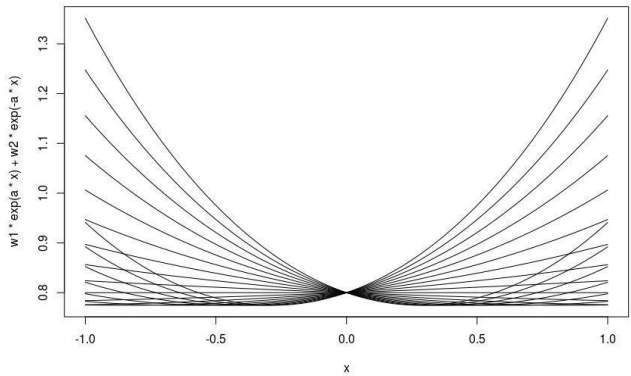

Figure 2: graph of equation 4 using values of a and b as obtained after fitting the non-linear least square curve. Coefficients a and b are the weights associated with current influence p(t) and historical influence p(-t) of journal.

The above figure shows the rate of change of influence of the journal with time. Time is represented on the x axis and the equation $w_1*exp(a*x)+w_2*exp(-a*x)$, which represents the influence at any time x, is represented on the y axis. The equation is a sum of two exponential curves, one with x as positive and one with x as a negative real number. At any point in time (x), either the positive or the negative part of the equation dominates. As we vary the value of $w_1$ and $w_2$, the above graph takes the following shape.

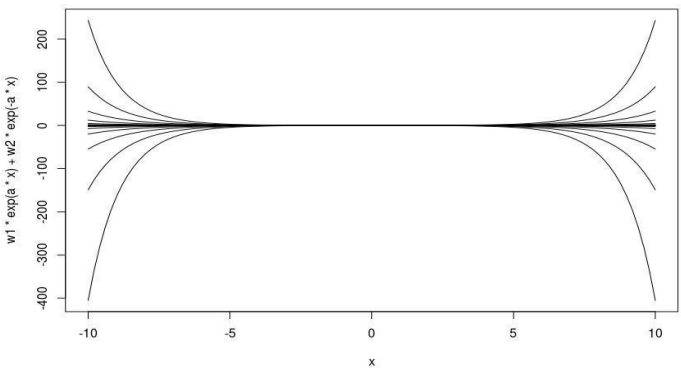

Figure 3: plot for equation (5) when w1<0 and w2>0 and when w1>0 and w2>0. Depending on the coefficients, the positive or the negative influence will dominate. In our case, since the influence is found to be positive, the parameter values responsible for negative influence are not considered for model tuning.

**Model Approximation: A Machine Learning Technique:**

The model equation and some initial values of the model are:

p'(t) =ap(t)+bp(-t)
p(0)=c; p'(o)=(a+b)c;

Approximately,
p'(t)=(p(t+h) - p(t))/h
and
p'(t)=(p(t) - p(t-h))/h

**Step 1.** At discrete time intervals, the values of p'(1), p'(2), p'(3), p'(4), p'(5) are computed by using the following equation.
p'(1)=ap(1)+bp(-1)

**Step 2**. Keeping the value of t as 1, in the differential equation (1) -
(p(1)-p(0))/1 = ap(1)+bp(-1)
=a[$Ae^{rt}$ + $Be^{-rt}$] + b[$Ae^{rt}$ + $Be^{-rt}$]
=(aA+bB)$e^r$ + (aA+bB)$e^{-r}$

**Step 3**: Compute the values of p'(¾), p'(½), p'(¼)

**Step 4:** The 4 coefficients a, b, A, B are measured by using the equations.

**Step 5:** Using non-linear least squares fitting technique, future rate-of-change of influence is predicted.

**MODEL IMPLICATIONS:**

From the results of this study, it can be established that the journal citations vary in a non-linear fashion. Initially, the journal citation score is usually less as the journal will have less popularity. This can be seen when we do a comparison of the citations of the editor v/s time and the influence of the journal v/s time. From figure 5, we observe that the citation of the editors is constant barring a few outliers. This will be utilized in future model modification. In the graph of influence v/s citations of editors (i.e. the reputation of the editors), we find that the influence of the journal remains more or less constant as the editors have significant reputation. This clearly shows that the initial influence of the journal depends on the influence of the editors who have started it.

The scibase and scientometric modeling effort is endorsed and supported by IEEE Computer Society Bangalore Chapter.

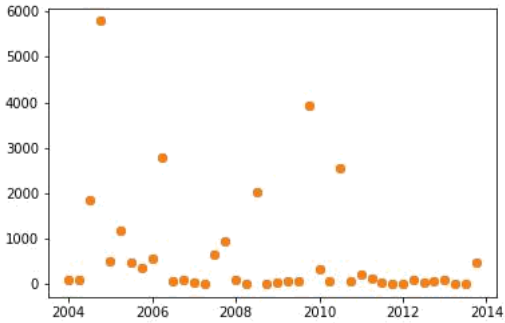

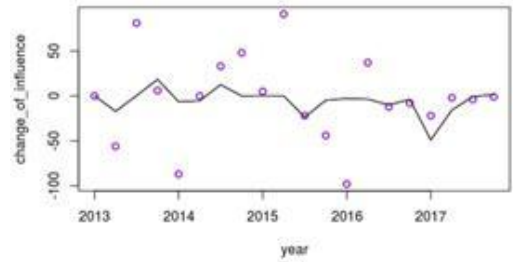

Figure 4: The above graph shows the citations of the editors from 2004-2014. We can observe that average citations within a span of every year do not vary greatly. Using this data we have fit the following graph of change in influence of journal VS editor's citations. This information is pertinent for inclusion in editorial influence in future modeling efforts where control variables play key roles in determining ascendancy in ASCOM's growth.

Figure 6: The above figure is a graph of editor's citations VS time. The citations are represented on the y-axis and time is represented on the x-axis. Since, the citations of the editors are more or less constant, the change in influence also shows a similar nature in spite of the initial irregularities. This information is pertinent for inclusion in editorial influence in future modeling efforts where control variables play key roles in determining ascendancy in ASCOM's growth.

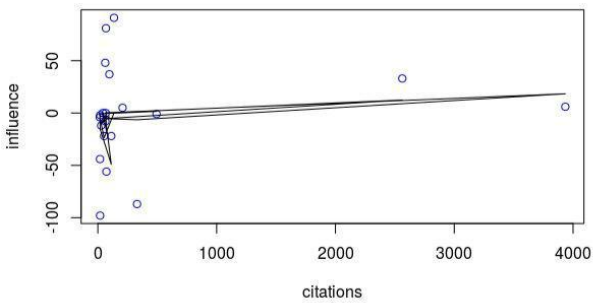

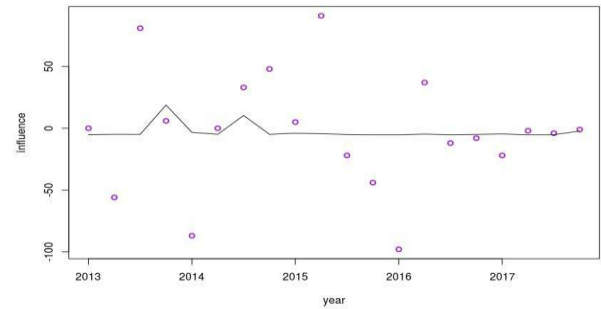

Figure 5: The above figure is a graph of change in influence of the journal against the editorial reputation (the super-star effect). We employed a least square fit between the two data sets and obtained reasonable degree of agreement between the two entities. This raises an important question: Does the editorial effort in rallying to consolidate a new journal go unnoticed? The impending investigation in to this complex question is beyond the scope of the manuscript but serves as a future research problem for us to explore. We intend to mitigate this by modifying our rudimentary model mentioned briefly in the section "Discussion and Future work".

Figure 7: (Non-linear least square curve). The change in influence over the span of 5 years shows small fluctuations but maintains an overall steady value.

The above graphs show that the rate of change in influence is more or less constant with time. There is an initial irregularity as the initial change in influence is directly related to the current influence. But with time the graph smoothed out because the other parameters such as citations and readership of the journal also begin to affect the rate of change of influence.

**DISCUSSION AND FUTURE WORK:**

We establish a model to study its effect on astronomy and computer science domains and analyze parameters that have contributed in building the reputation of ASCOM. The model explains the

The scibase and scientometric modeling effort is endorsed and supported by IEEE Computer Society Bangalore Chapter.

growth pattern of the journal well by capturing the intrinsic attributes and historical data. The time reversed model works as a mirror and helps carry over the good deeds of the past (quality of articles in niche areas and open problems solved by interdisciplinary efforts). Our model takes care of the "hereditary effects" and since the phenomenon of observing a journal in an emerging and interdisciplinary area is modeled as a function of spatial variables renders the system infinite degrees of freedom. Thus, the proposed model is robust provides better control over the system.

However, the data is limited since the journal is in publication for just over five years. Therefore the influence of historical data doesn't translate to overwhelming quantitative evidence in the way we liked it to. Additionally, we haven't considered implicit control variables which play important roles in the growth of any journal. These variables pose challenges to the modeling set up and without those, the scope is limited to empirical verification at minor scale.

When a journal is established by well-reputed editors, it receives a reasonable amount of attention because of the reputation of the editors. Put differently, when a journal is in its inception, the influence of the editorial board plays a significant role in building an initial impact of the journal. By 'influence' of editors, we mean the quality of work they publish in their domains. Apparently, the quality of work is portrayed by the impact of the published articles, citation count, h-index, h5-index, Research gate score, research projects, awards received and so on. Taking ASCOM as a case in point, its initial influence is represented as a function of present and past influence and later can be further embellished by the popularity of the editors who have laid its foundations (reputation of editors is gauged by the citation count).

We plan to render some model modifications to accommodate implicit control parameters such as publisher goodwill value and "start-up initiative" by editors. We define this initiative as the reputation of editors who shepherd the journal and act as strong attractors for quality submission from scholars across the globe. It is realistic to hypothesize that reputed scholars acting as editors add value and credibility to an emerging journal. This value however is extremely hard to quantify and therefore modeling such phenomenon is novel and imperative to understand the journal's growth pattern. We propose to present the model and the analytical solution, repeat the exercise of sections 3 and 4 and will discuss the implication of model modification in future.

The Time Reversed equation with the additive influence term (Publisher goodwill value) and the start-up initiative term can now be written as

$p'(t) = ap(t) + bp(-t) + \eta(t) + \theta(t)$

-where (t) is a constant additive term implying steady goodwill of the publishing house, Elsevier, in our case, θ(t), on the other hand represents Editors' reputation.

We shall also consider further modifications and additional constraints. These are as follows:

1. We shall assume (t) to be either linear or exponential and justify such considerations and rework the solutions to the new model.
2. We shall address the issue of quantifying the goodwill value. Is it a function of the percentage of accepted papers over time, a trend that accommodates a fixed number of accepted articles and the selection criteria of additional papers becomes more stringent? In the light of such assumptions, our model to extrapolate goodwill is formulated as:
    $(t) = \exp(-a) + \alpha \quad \alpha > 0, a > 0$

    -where $a$ is the percentage of articles accepted after the initial threshold of articles.

3. (t) is a control variable in the formulation and explanation of publisher goodwill and will be integrated with the modified model

Success of a journal depends mainly on editors and board members. The lead editor, editor-in-chief, lays strategic plans for the journal and makes crucial decisions related to manuscript submissions, adding new members to the board, forming a panel of reviewers etc. Evidently, a journal's success depends majorly on the editorial board constituency. Since editors are responsible for crucial decision and strategies, a journal's credibility is mapped with the credibility of editors particularly when the journal is in early stages of publication. Hence it is efficacious to spend time in discovering correlation between the journal's impact and the prestige of the editors who

The scibase and scientometric modeling effort is endorsed and supported by IEEE Computer Society Bangalore Chapter.

are managing the journal. The future work will identify relationship between a journal's growth (ASCOM) and the influence of editors to see if there exists positive correlation between the two factors. This can be proved by modelling the citations of editors of ASCOM by customizing the already implemented delay differential equations. System of delay differential equations (DDE) are explored in various science and biological domains, which uses not only the current initial value of variables but also the historical values to determine solution. Unlike the Ordinary differential equations, the derivative of solution y(t) in DDE depends on solution at present time as well as solution at previous time.

**REFERENCES**

[1] Gouri Ginde, Snehanshu Saha, Archana Mathur, Sukrit Venkatagiri, Sujith Vadakkepat, Anand Narasimhamurthy, B.S. Daya Sagar; ScientoBASE: A Framework and Model for Computing Scholastic Indicators of non-local influence of Journals via Native Data Acquisition algorithms; J. Scientometrics, Vol 107,# 1, pp 1-51, DOI 10.1007/s11192-016-2006-2

[2]. K. Bora, S. Saha, S. Agrawal, M. Safonova, S. Routh, A. Narasimhamurthy, CD-HPF: New habitability score via data analytic modeling, Astronomy and Computing, Volume 17, October 2016, Pages 129-143,ISSN2213-1337, http://dx.doi.org/10.1016/j.ascom.2016.08.001

[3]. Gouri Ginde, Snehanshu Saha, Chitra Balasubramaniam, Harsha R.S, Archana Mathur, B.S. Daya Sagar, Anand M N; Mining massive databases for computation of scholastic indices - Model and Quantify internationality and influence diffusion of peer-reviewed journals, Proceedings of the Fourth National Conference of Institute of Scientometrics, SIoT, August 2015 , pp 1-26

[4]. Saha, S., Jangid, N., Mathur, A., & Anand, M. N; DSRS: Estimation and Forecasting of Journal Influence in the Science and Technology Domain via a Lightweight Quantitative Approach, Collnet J. Scientometrics and Information Management, Vol 10, #1, pp 41-70 (Taylor and Francis), DOI:10.1080/09737766.2016.1177939

[5]. Saha, S., Sarkar, J., Dwivedi, A., Dwivedi, N., Narasimhamurthy, A., Mathur, A., & Roy, R. (2016). A novel revenue optimization model to address the operation and maintenance cost of a data center. Journal of Cloud Computing, Advances, Systems and Applications. DOI:10.1186/s13677-015-0050-8.

[6]. SciBase project: http://sahascibase.org/

[7]. Rahul Aedula, Yashasvi Madhukumar, Snehanshu Saha, Archana Mathur, Kakoli Bora and Surbhi Agrawal: L1 Norm SVD based Ranking Scheme: A Novel Method in Big Data Mining, Springer proceedings of the International Conference on Big data and Cloud Computing (2018).

[8]. Ginde G, Aedula R, Saha S, Mathur A, Dey S R, Sampatrao G S, Sagar B.:Big Data Acquisition, Preparation and Analysis using Apache Software Foundation Projects, Somani, A. (Ed.), Deka, G. (Ed.)., Big Data Analytics, New York: Chapman and Hall/CRC. (2017)

The scibase and scientometric modeling effort is endorsed and supported by IEEE Computer Society Bangalore Chapter.

The scibase and scientometric modeling effort is endorsed and supported by IEEE Computer Society Bangalore Chapter.